\documentclass[journal]{IEEEtran}
\usepackage{graphicx}
\usepackage{epstopdf}
\usepackage{cite}
\usepackage{algorithm}
\usepackage{algorithmic}
\usepackage{amsmath}
\usepackage{bm,comment,color,amssymb}
\usepackage{stfloats}
\usepackage{caption}
\usepackage{subcaption}

\usepackage[colorlinks]{hyperref}

\begin{document}

\title{Unlocking FAS-RIS Security Analysis with Block-Correlation Model}

\author{Jianchao Zheng, Xiazhi Lai, Tuo Wu, \\ Maged Elkashlan, Daniel Benevides da Costa,  \\ Chau Yuen, \emph{Fellow, IEEE}, Fumiyuki Adachi, \emph{Life Fellow}, \emph{IEEE} 

\thanks{ \emph{(Corresponding authors: Tuo Wu)}.
}
\thanks{ J. Zheng is with the School of Computer Science and Engineering, Huizhou University, Huizhou 516007, China (E-mail: $\rm zhengjch@hzu.edu.cn$). X. Lai is with the School of Computer Science, Guangdong University of Education, Guangzhou, Guangdong, China (E-mail: $\rm xzlai@outlook.com$). T. Wu and C. Yuen are with the School of Electrical and Electronic Engineering, Nanyang Technological University, 639798, Singapore (E-mail: $\rm \{tuo.wu, chau.yuen\}@ntu.edu.sg$). M. Elkashlan is with the School of Electronic Engineering and Computer Science at Queen Mary University of London, London E1 4NS, U.K. (E-mail: $\rm maged.elkashlan@qmul.ac.uk$). D. B. da Costa is with the Interdisciplinary Research Center for Communication Systems and Sensing (IRC-CSS), Department of Electrical Engineering, King Fahd University of Petroleum $\&$ Minerals (KFUPM), Dhahran 31261, Saudi Arabia (E-mail:$\rm danielbcosta@ieee.org$). F. Adachi is with the International Research Institute of Disaster Science (IRIDeS), Tohoku University, Sendai, Japan (E-mail: $\rm adachi@ecei.tohoku.ac.jp$).
}
} 

\markboth{}
{Zheng \MakeLowercase{\textit{et al.}}:  Unlocking FAS-RIS  Security Analysis with Block-Correlation Model}

\maketitle

\begin{abstract}
In this letter, we investigate the security of fluid antenna system (FAS)-reconfigurable intelligent surfaces (RIS) communication systems. The base station (BS) employs a single fixed-position antenna, while both the legitimate receiver and the eavesdropper are equipped with fluid antennas.  By utilizing the block-correlation model and the central limit theorem (CLT), we derive approximate expressions for the average secrecy capacity and secrecy outage probability (SOP). Our analysis, validated by simulation results, demonstrates the effectiveness of the block-correlation model in accurately assessing the security performance. Moreover,   simulation results reveal that FAS-RIS system  significantly outperforms other systems in terms of security, further underscoring its potential in secure communication applications.
\end{abstract}

\begin{IEEEkeywords}
Fluid antenna system (FAS), reconfigurable intelligent surfaces (RIS), average secrecy capacity, secrecy outage probability (SOP).
\end{IEEEkeywords}
\IEEEpeerreviewmaketitle

\section{Introduction}
\IEEEPARstart{F}{luid} antenna systems (FAS) have emerged as a key technology for next-generation wireless communications, offering enhanced flexibility and adaptability compared to traditional antenna systems \cite{FAS20}. In practice, FAS can be implemented using either pixel-based structures \cite{Rodrigo14} or liquid metal-based designs \cite{Huang21}, both of which enable dynamic reconfiguration to optimize performance. The selection of the optimal antenna port can be achieved through learning-based techniques that exploit spatial correlation to maximize signal quality \cite{Chai22}.

Research on FAS has gained significant attention, covering a wide range of topics, including maximum ratio combining for improved signal-to-noise ratio \cite{XLai23}, short-packet communications for ultra-reliable low-latency scenarios \cite{XLai23}, proactive monitoring for enhanced network reliability \cite{YaoJ24}, and efficient channel estimation techniques \cite{HXu24}. These studies highlight the potential of FAS in addressing various challenges of modern wireless communication systems, such as improving spectral efficiency, reliability, and adaptability in dynamic environments.

In FAS-assisted communication systems, adjusting the fluid antennas (FAs) to their optimal locations is crucial for selecting the most favorable channels, thereby directly impacting communication performance \cite{Alvim-2023, Dai-2023, Xu-2024}. However, this adjustment process often involves significant computational complexity and operational overhead, which can hinder practical deployment. To address these issues, reconfigurable intelligent surfaces (RIS) \cite{KZhi221, TWu1, TWu2} have emerged as a promising technology. By leveraging their phase adjustment capabilities, RIS can mitigate the impact of suboptimal FA positioning by providing enhanced control over the channel environment, effectively compensating for potential performance losses.

These features make RIS an attractive addition to FAS-assisted systems, enabling more flexible and efficient channel optimization. As a result, FAS-RIS communication systems have gained substantial research interest, with numerous studies focusing on different aspects, including outage analysis \cite{LaiX242, Ghadi2024} and performance optimization \cite{YaoJ242, YaoJ243}. This combination of technologies holds great potential for enhancing the reliability and efficiency of next-generation wireless communication networks.
 
However, FAS-RIS communication systems also face security challenges that must be addressed. While the authors in \cite{Ghadi2024}  {have} considered this issue, their approach requires calculating the cumulative distribution function (CDF) of an $N$-dimensional Gaussian distribution { ($N$ is the number of ports)}, involving numerically integrating over $N$ dimensions, which is computationally complex. To address this challenge, we introduce an efficient approximation model, the block-correlation model \cite{Espinosa24}, which simplifies the channel correlation coefficients uniformly to facilitate analysis, significantly reducing the computational complexity.

In this letter, we investigate a FAS-RIS communication system comprising a base station (BS), a legitimate receiver, and an eavesdropper. Both the legitimate receiver and the eavesdropper are equipped with flexible arrays (FAs), enabling dynamic repositioning among $N$ preset locations (or ports). By applying the block-correlation model, we derive approximate expressions for the average secrecy capacity and the secrecy outage probability (SOP). The simulation results demonstrate that the proposed FAS-RIS system significantly outperforms existing systems in terms of security.

\section{System Model} 
Consider a FAS-RIS secure communication system comprising a BS equipped with a single fixed-position antenna (FPA), one legitimate receiver, and one eavesdropper, both equipped with FA, along with a  RIS containing \(2M\) reflecting elements. The FAs can switch among \(N\) ports within a linear space of \(W\lambda\), where \(W\) is a normalization parameter and \(\lambda\) represents the wavelength. In this system, the legitimate receiver is the legitimate user, while the eavesdropper  attempts to intercept the signals intended for the legitimate receiver. The BS transmits signals to the legitimate receiver with the aid of RIS, as the direct channels between the BS and {the legitimate receiver and the eavesdropper} are obstructed by physical barriers such as buildings.  To simplify the analysis, the RIS is divided into two zones, each containing \(M\) reflecting elements \cite{BariahL21,LiS22}. Each zone provides a distinct phase shift to redirect the signals toward a specific user. The first zone is designated to reflect signals from the BS to the legitimate receiver, while the second zone directs signals to the eavesdropper.

\subsection{Communication Model}
The BS transmits the signals to {the legitimate receiver} solely through reflections from the RIS. Consequently, the received signal at the $k$-th FAS port for the legitimate receiver and eavesdropper is expressed as 
\begin{align}
y_k^{(x)}=\sqrt{P}\sum_{m=1}^M h_m^{(x)} v_{m,k}^{(x)} e^{-2i\pi\theta_m^{(x)}}s+n_k^{(x)},
\end{align}
where $x=\{r, e\}$ represent the legitimate receiver or eavesdropper, respectively. {Here, the Rayleigh fading channel is considered, which is suitable for non-light-of-sight scenarios}, $h_m^{(x)}\sim \mathcal{CN}(0, \epsilon_1^{(x)})$ denotes the channel coefficient from the BS to the $m$-th RIS element in the first or second zone, and $v_{m,k}^{(x)}\sim \mathcal{CN}(0, \epsilon_2^{(x)})$ represents the channel coefficient from the $m$-th RIS element in the first or second zone to the $k$-th FAS port for the legitimate receiver or eavesdropper. The transmit signal is denoted by $s\sim \mathcal{CN}(0, 1)$, the transmit power by $P$, and $n_k^{(x)}\sim \mathcal{CN}(0, \sigma_{(x)}^2)$ represents the additive white Gaussian noise (AWGN). The term $\theta_m^{(x)}$ denotes the reflection phase of the $m$-th element in the first or second zone of the RIS. With full knowledge of the phases of $h_m^{(x)}$  and  $v_{m,k}^{(x)}$ \footnote[1]{ {Channel estimation techniques can be applied here for obtaining channel state information (CSI) \cite{HXu24,FengH24}. }}, the RIS adjusts $\theta_m^{(x)}$ to maximize the channel gain, specifically by setting $\theta^{(x)}=-\text{arg}\left(h_m^{(x)} v_{m,k}^{(x)}\right)$. Under this assumption, the cascaded channel {\cite{ZH20}} parameter at the $n$-th FAS port of the legitimate receiver and eavesdropper can be written as 
\begin{align}  \label{aa1}
	\gamma_k^{(x)}= \sum_{m=1}^M|h_m^{(x)}||v_{m,k}^{(x)}|.
\end{align}

{Both} $|h_m^{(x)}|$, and $|v_{m,k}^{(x)}|$ follow a Rayleigh distribution, the expected value and variance of the product $|h_m^{(x)}||v_{m,k}^{(x)}|$ are given by
\begin{align}   
	&\mathbf{E}\left(|h_m^{(x)}||v_{m,k}^{(x)}|\right)\nonumber\\
	&=\mathbf{E}\left(|h_m^{(x)}|\right)\mathbf{E}\left(|v_{m,k}^{(x)}|\right)=\pi\sqrt{\epsilon_1^{(x)}\epsilon_2^{(x)}}/4,\nonumber\\
	&\mathbf{Var}\left(|h_m^{(x)}||v_{m,k}^{(x)}|\right)\nonumber\\
	&=\mathbf{E}\left(|h_m^{(x)}|^2|v_{m,k}^{(x)}|^2\right)-\left(\mathbf{E}\left(|h_m^{(x)}||v_{m,k}^{(x)}|\right)\right)^2\nonumber\\
		&=\epsilon_1^{(x)}\epsilon_2^{(x)}(1-\pi^2/16).
\end{align} 
Consequently, the maximum channel gain can be obtained by selecting the optimal FAS port for signal receiving. Specifically, the channel parameter of the optimal port is given by
\begin{align}
	\gamma_{(x)}^*=\mathop{\text{max}}\limits_{k=1,\cdots, N} \gamma_k^{(x)}.
\end{align}
\subsection{FAS Channel Correlation Model}
For the {channel correlation} between two different ports of the FAS, the 3D Clarke's model \cite{Aulin79} \footnote[2]{ {The spatial correlation in Rayleigh channels depends on the antenna radiation pattern and the distribution of angles of arrival from the propagation environment. Isotropic antennas and propagation are commonly assumed, leading to the 3D Clarke’s model. Jakes’s correlation is often used for 1D fluid antennas, but it is not suitable for planar FAS as it assumes a 2D propagation environment \cite{Espinosa24}.}} is adopted. Specifically, the correlation coefficient between ports $\varsigma$ and $\tau$ is modeled by 
\begin{align}
	g^{(x)}(\varsigma,\tau)=\text{sinc}\left(\frac{2\pi (\varsigma-\tau) W}{N-1}\right),
\end{align}
where $\mathrm{sinc}(x)=\frac{\sin(x)}{x}$. Note that $\mathrm{sinc}(x)$ is an odd function, which indicates that the correlation coefficient matrix is a Toeplitz matrix as 
\begin{equation}
	\mathbf{\Sigma}^{(x)}\in\mathbb{R}^{N\times N}=
	\begin{bmatrix}\label{q5}
		g_{1,1}^{(x)} & g_{1,2}^{(x)} & \dots & g_{1,N}^{(x)}\\
		g_{1,2}^{(x)} & g_{1,1}^{(x)} & \dots & g_{1,N-1}^{(x)}\\
		\vdots &  \ddots & \vdots \\
		g_{1,N}^{(x)} & g_{1,N-1}^{(x)} & \dots & g_{1,1}^{(x)}
	\end{bmatrix}.
\end{equation}

\section{Average secrecy capacity and SOP analysis}
In this section, we will derive the theoretical  {approximate expressions} for the average secrecy capacity and secrecy outage probability (SOP). First, by using the optimal channel gain, the mutual information rates for the signal $s$ at the legitimate receiver and eavesdropper can be expressed as
\begin{align}
	R_r=&\frac{1}{2}\log_2\left(1+\frac{P}{\sigma_r^2}\left(\gamma_r^*\right)^2\right),\\
	R_e=&\frac{1}{2}\log_2\left(1+\frac{P}{\sigma_e^2}\left(\gamma_e^*\right)^2\right).
\end{align}
The secrecy capacity, which represents the difference between the mutual information rates of the legitimate user and the eavesdropper, is then given by
\begin{align}  
	C_s=\left[R_r-R_e\right]^+,\label{q6}
\end{align}
where $[a]^+=\max\{a,0\}$. Finally, for a target secrecy rate $R_s$,  the {SOP} can be calculated as
\begin{align}\label{q9}
	P_{\mbox{sop}}=\Pr\left(C_s<R_s\right).
\end{align}
\subsection{Average secrecy capacity}
First, the theoretical analysis of the average secrecy capacity is calculated as 
\begin{align} \label{ab2}
\bar{C_s}=\mathbf{E}\left(C_s\right)=&\int_0^\infty\int_0^{\frac{\sigma_e}{\sigma_r}x} C_s f_{\gamma_r^*,\gamma_e^*}(x,y)dydx,
\end{align}
where $f_{\gamma_r^*,\gamma_e^*}(x,y)=f_{\gamma_r^*}(x)f_{\gamma_e^*}(y)$ is the joint PDF of $f_{\gamma_r^*}(x)$ and $f_{\gamma_e^*}(y)$. Combining Eq. \eqref{q6} with  Eq. \eqref{ab2}, we have
\begin{align}
\bar{C_s}=&\frac{1}{2}\int_0^\infty\int_0^{\frac{\sigma_e}{\sigma_r}x} \left(\log_2\left(1+\frac{P}{\sigma_r^2}x^2\right)-\log_2\left(1+\frac{P}{\sigma_e^2}y^2\right)\right)\nonumber\\
&\times f_{\gamma_r^*}(x)f_{\gamma_e^*}(y)dydx\nonumber\\
=&C_s^1-C_s^2,
\end{align}
where 
\begin{align}
C_s^1=&\frac{1}{2}\int_0^\infty\int_0^{\frac{\sigma_e}{\sigma_r}x} \log_2\left(1+\frac{P}{\sigma_r^2}x^2\right) f_{\gamma_r^*}(x)f_{\gamma_e^*}(y)dxdy\nonumber\\
=&\frac{1}{2}\int_0^\infty\log_2\left(1+\frac{P}{\sigma_r^2}x^2\right) f_{\gamma_r^*}(x)F_{\gamma_e^*}\left(\frac{\sigma_e}{\sigma_r}x\right)dx,
\end{align}
and 
\begin{align}
C_s^2=&\frac{1}{2}\int_0^\infty\int_0^{\frac{\sigma_e}{\sigma_r}x}\log_2\left(1+\frac{P}{\sigma_e^2}y^2\right) f_{\gamma_r^*}(x)f_{\gamma_e^*}(y)dydx.
\end{align}
Consequently, we will derive the PDF of $ f_{\gamma_r^*}(x)$ and $ f_{\gamma_e^*}(y)$. First, to efficiently capture the significant eigenvalues, we employ  the block-correlation approximation model \cite{Espinosa24} to simplify  the representation of the  {Eq. \eqref{q5}}. Accordingly, the correlation coefficient matrix can be formulated as
\begin{equation}
  \hat{ \mathbf{\Sigma}}^{(x)}\in\mathbb{R}^{N\times N}=
	\begin{bmatrix}
	\mathbf{A}_1^{(x)} &\mathbf{0} &\mathbf{0} & \dots &\mathbf{0}\\
	\mathbf{0} &\mathbf{A}_2^{(x)} &\mathbf{0}& \dots & \mathbf{0}\\
	 \vdots &  \ddots & \vdots \\
	 \mathbf{0}& \mathbf{0} & \mathbf{0}& \dots & \mathbf{A}_B^{(x)}
	\end{bmatrix},
\end{equation}
where 
\begin{equation}
 \mathbf{A}_b^{(x)}\in\mathbb{R}^{L_b^{(x)}\times L_b^{(x)}}=
	\begin{bmatrix}
	1 &\mu^{(x)} &\mu^{(x)} & \dots &\mu^{(x)}\\
	\mu^{(x)} &1&\mu^{(x)}& \dots &\mu^{(x)}\\
	 \vdots &  \ddots & \vdots \\
	 \mu^{(x)}& \mu^{(x)} & \mu^{(x)}& \dots & 1
	\end{bmatrix},
\end{equation}
$\mu^{(x)}$ is a number close to 1, and $\sum_{L_b^{(x)}=1}^B L_b=N$, $B=|\it{S(\lambda^{(x)}_{th})}|$ is the cardinal number of  $\it{S(\lambda^{(x)}_{th})}$. And  $\it{S(\lambda^{(x)}_{th})}=\left\{\lambda_n^{(x)}|\lambda_n^{(x)}\geq\lambda_{th}, n=1,\cdots,N\right\}$, where $\lambda_{th}$ is a tiny number to ensure that enough eigenvalues have been included in $\it{S(\lambda^{(x)}_{th})}$, and it can be dynamically adjusted. $L_b$ is determined based on
\begin{equation}
\mathop {\arg\min}\limits_{L_1,\cdots, L_B} \text{dist}( \mathbf{\Sigma}^{(x)}, \hat{ \mathbf{\Sigma}}^{(x)}),
\end{equation} 
where dist(·) is a distance metric between two matrices, which is determined by the difference of their eigenvalues and the detailed procedure can be found in \cite{Espinosa24,LaiX242}. To gain more insight, we consider the setting that $\mu^{(x)}=1$, i.e., perfect correlation between the ports within each block. In such case, each block can be regarded as a single antenna, and the FAS directly approximated by a collection of $B$ independent antennas.

Based on the block-correlation model described above, and assuming that the number of RIS elements $M$ is sufficiently large, the central limit theorem (CLT) approximation can be applied \cite{LaiX242}.  Under this approximation, $\mathbf{\Gamma}^{(x)}=[\gamma_1^{(x)},\cdots,\gamma_N^{(x)}]$ can be approximated by $\hat{\mathbf{\Gamma}}^{(x)}=[\hat{\gamma}_1^{(x)},\cdots,\hat{\gamma}_N^{(x)}]$. Consequently, by applying the Pearson correlation coefficient calculation, the correlation coefficient matrix of $\hat{\mathbf{\Gamma}}^{(x)}$ is given by
\begin{align}
	&\bar{ \mathbf{\Sigma}}^{(x)}\in\mathbb{R}^{N\times N}\nonumber\\
	&=\begin{bmatrix}
		\mathbf{D}_1^{(x)} &\mathbf{C} &\mathbf{C} & \dots &\mathbf{C}\\
		\mathbf{C} &\mathbf{D}_2^{(x)} &\mathbf{C}& \dots & \mathbf{C}\\
		\vdots &  \ddots & \vdots \\
		\mathbf{C}&\mathbf{C} & \mathbf{C}& \dots & \mathbf{D}_B^{(x)}
	\end{bmatrix},
\end{align}
where $\mathbf{D}_b^{(x)}\in\mathbb{R}^{L_b^{(x)}\times L_b^{(x)}}=\mathbf{1}_{N\times N}$.  
Additionally, all elements in submatrix within $\mathbf{C}$ are given as $\rho_0=\frac{\pi(4-\pi)}{16-\pi^2}$.
\begin{align}
\rho_0=\frac{\pi(4-\pi)}{16-\pi^2}.
\end{align}
Thanks to the block-correlation model, where the correlation coefficient within each block is a constant, the analysis becomes tractable for FAS-RIS communication. Specifically, the expression for $\bar{\gamma}^{(x)}_k$ is given by
\begin{align}
\bar{\gamma}^{(x)}_k=\sqrt{1-\rho^{(x)}_0}d^{(x)}_k+\sqrt{\rho^{(x)}_0}d^{(x)}_0+E_{\gamma_{(x)}},
\end{align}
where $d^{(x)}_k\sim \mathcal{N}(0, V_{\gamma_{(x)}})$ for $k=0,1,2,\cdots, B$.  In addition, we have $E_{\gamma_{(x)}}= \frac{M\pi\sqrt{\epsilon_1^{(x)}\epsilon_2^{(x)}}}{4}$, $V_{\gamma_{(x)}}= M\epsilon_1^{(x)}\epsilon_2^{(x)}\left(1-\frac{\pi^2}{16}\right)$. 
Building upon this, the CDF of $\bar{\gamma}_{(x)}^*$ can be formulated as \cite{LaiX242}
\begin{align}
&F_{\bar{\gamma}_{(x)}^*}(y_{(x)})\nonumber\\
&\approx  \frac{H\pi}{U_l}\sum_{l=1}^U \frac{1}{2^B}\left[1+\text{erf}\left(\frac{y_{(x)}-E_{\gamma_{(x)}}-\rho^{(x)}_0\tau}{\sqrt{2V_{\gamma_{(x)}}(1-\rho^{(x)}_0)}}\right)\right]^B\nonumber\\
&\quad\times\sqrt{\frac{1-q_l^2}{2\pi V_{\gamma_{(x)}}}}e^{-\frac{\tau^2}{2V_{\gamma_{(x)}}}},
\end{align}
where $H$ is a huge number, $U_l$ is a parameter controlling the accuracy-complexity tradeoff, $\tau=\frac{Hq_l+H}{2}$, $q_l=\cos\left(\frac{(2l-1)\pi}{2U}\right)$. From the CDF, the PDF of $\bar{\gamma}^*$ is given by
\begin{align}
&f_{\bar{\gamma}_{(x)}^*}(y_{(x)})\nonumber\\
&\approx  \frac{H}{UV_{\gamma}}\sum_{l=1}^U \frac{B}{2^B}\left[1+\text{erf}\left(\frac{y_{(x)}-E_{\gamma_{(x)}}-\rho^{(x)}_0\tau}{\sqrt{2V_{\gamma_{(x)}}(1-\rho^{(x)}_0)}}\right)\right]^{B-1}\nonumber\\
&\quad\times\sqrt{\frac{1-q_l^2}{1-\rho^{(x)}_0}}e^{-\frac{y_{(x)}-E_{\gamma}-\rho_0\tau}{2V_{\gamma_{(x)}}(1-\rho^{(x)}_0)}-\frac{\tau^2}{2V_{\gamma_{(x)}}}}.
\end{align}
Finally, the average secrecy capacity can be derived as
\begin{align}
\bar{C_s}\approx&\frac{H\pi}{4 U_p}\sum_{p=1}^{U_p}\sqrt{1-t_p^2}\log_2\left(1+\frac{P}{\sigma_r^2}\beta^2\right) f_{\bar{\gamma}_r^*}(\beta)F_{\bar{\gamma}_e^*}\left(\frac{\sigma_e}{\sigma_r}\beta\right)\nonumber\\
&-\frac{H\pi^2\sigma_e}{8 U_p U_l \sigma_r}\sum_{p=1}^{U_p}\sqrt{1-t_p^2}\sum_{l=1}^{U_l}\sqrt{1-q_l^2}\log_2\left(1+\frac{P}{\sigma_e^2}\chi^2\right)\nonumber\\
&\times f_{\bar{\gamma}_r^*}(\beta)f_{\bar{\gamma}_e^*}(\chi),
\end{align}
where the Gauss-Chebyshev integral is applied in the final step, $U_p$ is the accuracy-complexity trade-off parameter, $t_p=\cos\left(\frac{(2p-1)\pi}{2U_p}\right)$, $\beta=\frac{Ht_p+H}{2}$, and $\chi=\frac{\sigma_e}{2\sigma_r}\beta(ql+1)$.

 {\subsection{SOP}}
In this subsection, we analyze the SOP of the considered system. {For Eq. \eqref{q9}, $R_s$ is defined as the target secrecy rate.} Accordingly, the SOP is defined as $P_{\text{SOP}}=\Pr\left(C_s<R_s\right)$. Substituting  Eq. \eqref{q6} into  this expression, the SOP can be further derived as
\begin{align}
 &P_{\text{SOP}} \nonumber\\
	\approx&\frac{H\pi}{2U_p}\sum_{p=1}^{U_p}\sqrt{1-t_p^2}F_{\bar{\gamma}_r^*}\left( {\sqrt{\frac{\sigma_r^2}{P}\left(2^{2R_s}\left(1+\frac{P}{\sigma_e^2}\beta^2\right)-1\right)}}\right)\nonumber\\
	&\times f_{\bar{\gamma}_e^*}(\beta),
\end{align}
where the Gauss-Chebyshev integral is applied in the approximation of the final step.

 \begin{figure}[t]
 	\centering
 	\includegraphics[width=0.98\linewidth]{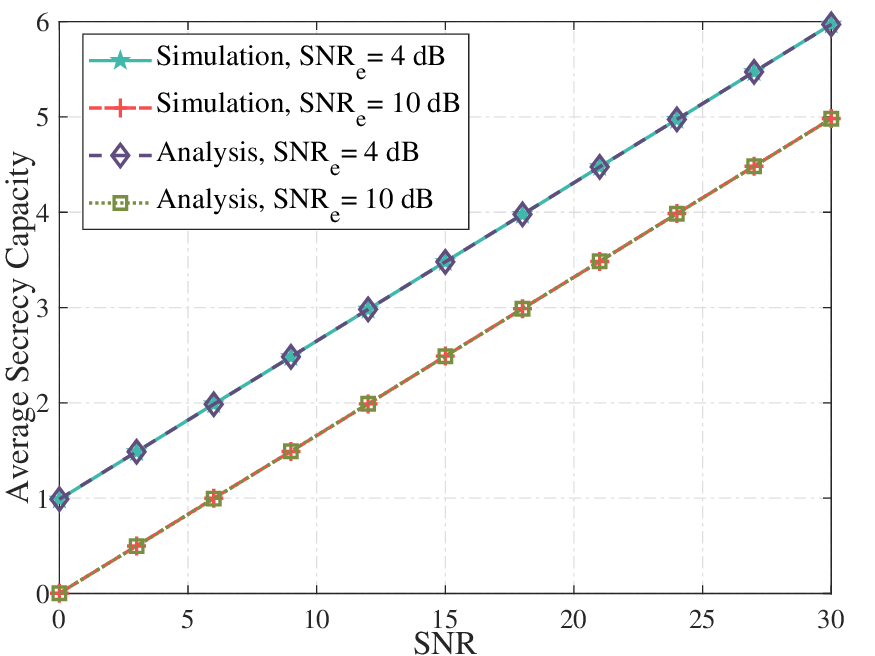}
 	\caption{Average secrecy capacity versus SNR.}\label{fig1} 
 \end{figure}
 \begin{figure}[t]
 	\centering
 	\includegraphics[width=\linewidth]{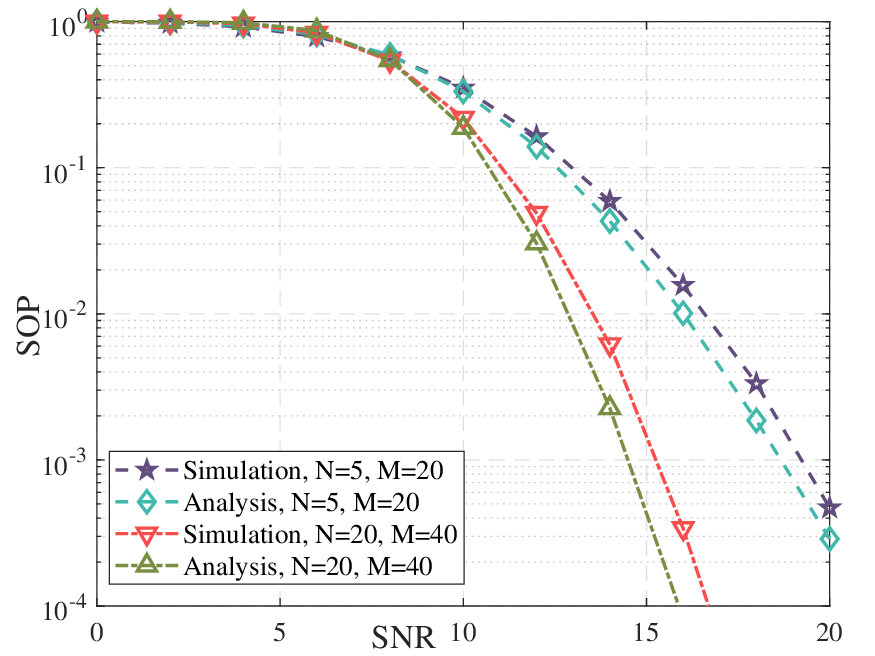}
 	\caption{SOP v.s. SNR.}\label{fig3}
 \end{figure}

\section{Numerical Results} 

In this section, we present numerical results to validate the accuracy of the analytically derived average secrecy capacity and SOP. {And the RIS is considered to adjust $\theta_m^{(x)}$ to maximize the channel gain Eq. \eqref{aa1} for the legitimate user.} For our simulations, we adopt the following parameters: $\epsilon_1^r =\epsilon_2^r=\epsilon_1^e  = \frac{1}{2}$, $\epsilon_2^e  = \frac{1}{20}$, and $W = 5$, {$R_s=2$ bit/s/Hz}.

In Fig.~\ref{fig1}, the average secrecy capacity is shown for a fixed $N=20$ and $M = 40$, with either $\text{SNR}_e = 4$ dB or $\text{SNR}_e = 10$ dB, where $\text{SNR}_e$ represents the signal-to-noise ratio (SNR) at the eavesdropper. ``Simulation'' and ``Analysis'' represent the simulated and  theoretical average secrecy capacity. From Fig.~\ref{fig1}, the average secrecy capacity increases significantly with the rise in SNR.The close agreement between the theoretical and simulated results validates the accuracy of the analytical block-correlation model. Moreover, the block-correlation model effectively avoids the heavy computation involved in the CDF of multi-dimensional Gaussian distributions, simplifying the performance analysis of FAS-RIS systems.

Fig.~\ref{fig3} depicts the relationship between the SOP and SNR. We have set the SNR for legitimate receiver $\text{SNR}_r = 10$ dB and for eavesdropper $\text{SNR}_e = 4$ dB, with $N = 5, M=20$ or $N = 20, M=40$.  The results  in Fig.~\ref{fig3}, demonstrate the consistency between the theoretical analysis and the simulated outcomes for the SOP, validating the robustness of the block-correlation approximation model. It can also be observed that as the values of $M$ or $N$ increase, the outage probability decreases, which is due to the enhanced diversity achieved by the FAS-enabled users with larger $M$ or $N$.

\begin{figure}[t]
	\centering
	\includegraphics[width=\linewidth]{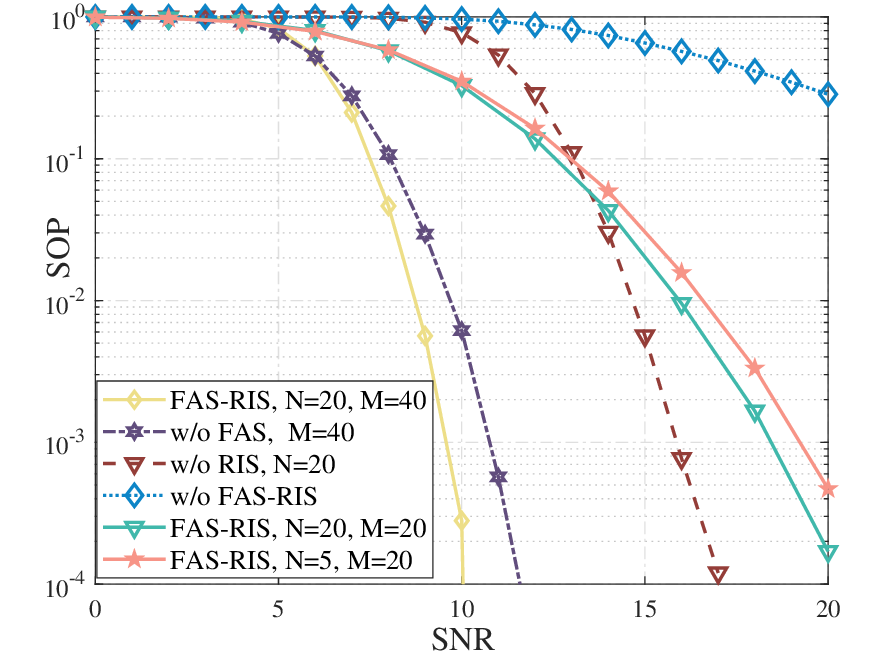}
	\caption{SOP v.s. SNR for different scenarios.}\label{fig4}
\end{figure}
Fig.~\ref{fig4} illustrates the relationship between the SOP and SNR for various configurations, assuming $N=20$ and $M=40$. Fig.~\ref{fig4} compares the performance of FAS-RIS systems against scenarios without RIS, without FAS, and without FAS-RIS. The curves clearly demonstrate that the FAS-RIS configuration significantly outperforms the others, especially in the presence of RIS, which plays a crucial role in enhancing security. Moreover, the effectiveness of FAS is also evident, underscoring the importance of both FAS and RIS in improving communication security. Additionally, the graph shows that increasing the number of FAS ports and RIS reflecting elements further enhances the security performance, reinforcing the critical role of these parameters in FAS-RIS systems.

\textbf{Remark 1}: \textit{Our results highlight the importance of both FAS and RIS in enhancing system security. The absence of either component leads to notable performance degradation, while their integration significantly improves communication security. Increasing the number of FAS ports and RIS reflecting elements further boosts security, underscoring the synergistic effect of FAS and RIS in FAS-RIS systems. These insights provide valuable guidance for the design of secure communication systems in future wireless networks.}

\begin{figure}[t]
	\centering
	\includegraphics[width=0.96\linewidth]{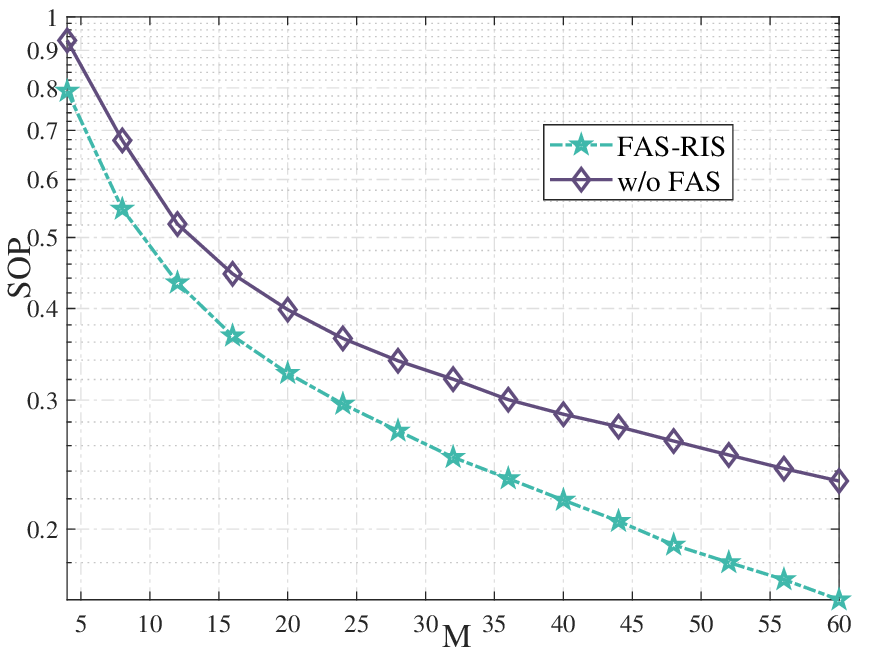}
	\caption{SOP v.s. M.}\label{fig5}
\end{figure}
{Fig.~\ref{fig5} illustrates the relationship between the SOP and $M$, assuming $N=20$. The curves clearly demonstrate that the FAS-RIS systems significantly outperforms the without FAS configuration. Moreover, the effectiveness of FAS is more evident when the number of RIS reflecting elements become larger, which further demonstrate the importance of FAS.}

\textbf{Remark 2}: \textit{The results show that FAS significantly enhances system performance, especially as the number of RIS reflecting elements increases. The pronounced difference between the FAS-RIS system and the configuration without FAS highlights the critical importance of FAS in improving security.}

\section{Conclusion}
In this letter, we analyzed a FAS-RIS secure communication systems. By employing the block-correlation approximation model, we derived {the approximate expressions} for the average secrecy capacity and SOP of the FAS-RIS secure communication system. {The simulation results revealed that FAS-RIS system significantly outperforms other system.}


\begin{thebibliography}{1}
\bibitem{FAS20}
K.-K. Wong, A. Shojaeifard, K.-F. Tong and Y. Zhang, ``Performance limits of fluid antenna systems," \emph{IEEE Commun. Letters}, vol. 24, no. 11, pp. 2469--2472, Nov. 2020.
 
\bibitem{Rodrigo14}
D. Rodrigo, B. A. Cetiner, and L. Jofre, ``Frequency, radiation pattern and polarization reconfigurable antenna using a parasitic pixel layer," \emph{IEEE Trans. Antennas \& Propag.}, vol. 62, no. 6, pp. 3422--3427, Jun. 2014.
\bibitem{Huang21}
Y. Huang, L. Xing, C. Song, S. Wang, and F. Elhouni, ``Liquid antennas: Past, present and future," \emph{IEEE Open J. Antennas \& Propag.}, vol. 2, pp. 473--487, Mar. 2021.

\bibitem{Chai22}
Z. Chai, K.-K. Wong, K.-F. Tong, Y. Chen, and Y. Zhang, ``Port selection for fluid antenna systems," \emph{IEEE Commun. Lett.}, vol. 26, no. 5, pp. 1180--1184, May 2022.

\bibitem{XLai23}
X. Lai {\em et al.}, ``On performance of fluid antenna system using maximum ratio combining," \emph{IEEE Commun. Lett.}, vol. 28, no. 2, pp. 402--406, Feb. 2024.

\bibitem{JZheng24}
J. Zheng {\em et al.}, ``FAS-assisted NOMA short-packet communication systems," \emph{IEEE Trans. Veh. Technol. }, vol. 73, no. 7, pp. 10732--10737, Jul.  2024.

\bibitem{YaoJ24} 
J. Yao et al., ``Proactive monitoring via jamming in fluid antenna systems," \emph{IEEE Commun. Lett.}, vol. 28, no. 7, pp. 1698--1702, Jul. 2024. 
 
 
 \bibitem{HXu24}
 H. Xu {\em et al.}, ``Channel Estimation for FAS-Assisted Multiuser mmWave Systems," \emph{IEEE Commun. Lett.}, vol. 28, no. 3, pp. 632-636, Mar. 2024.  
 
\bibitem{Alvim-2023}
P. D. Alvim {\em et al.}, ``On the performance of fluid antennas systems under $\alpha$-$\mu$ fading channels,'' {\em IEEE Wireless Commun. Lett.}, vol. 13, no. 1, pp. 108--112, Jan. 2024.

 
\bibitem{Dai-2023}
Z. Zhang, J. Zhu, L. Dai, and R. W. Heath Jr, ``Successive Bayesian reconstructor for channel estimation in fluid antenna systems,'' arXiv preprint \url{arXiv:2312.06551v3}, 2024.
 
\bibitem{Xu-2024}
Y. Chen, and T. Xu, ``Fluid antenna index modulation communications,'' {\em IEEE Wireless Commun. Lett.}, vol. 13, no. 4, pp. 1203--1207, Apr. 2024.

\bibitem{KZhi221}
K. Zhi, C. Pan, H. Ren, and K. Wang, ``Ergodic rate analysis of reconfigurable intelligent surface-aided massive MIMO systems with ZF detectors,'' \emph{IEEE Commun. Lett.}, vol. 26, no. 2, pp. 264--268, Feb. 2022.

\bibitem{TWu1}
T. Wu {\em et al.}, ``Fingerprint-based mmWave positioning system aided by reconfigurable intelligent surface,"  \emph{IEEE Wireless Commun. Lett.}, vol. 12, no. 8, pp. 1379--1383, Aug. 2023.
 \bibitem{TWu2}
T. Wu {\em et al.}, ``Exploit High-Dimensional RIS Information to Localization: What Is the Impact of Faulty Element?,"  \emph{IEEE J. Sel. Areas Commun.}, early access, \url{doi:10.1109/JSAC.2024.3414582}, 2024.


\bibitem{LaiX242} 
X. Lai, J. Yao, K. Zhi, T. Wu, D. Morales-Jimenez and  K. K. Wong,  ``FAS-RIS: A Block-Correlation Model Analysis," to be published.

\bibitem{Ghadi2024}
F. R. Ghadi {\em et al.}, ``Secrecy performance analysis of RIS-Aided fluid antenna systems," arXiv preprint \url{arXiv:2408.14969}, Aug. 2024.

\bibitem{YaoJ242} 
J. Yao, {\em et al.}, ``FAS-RIS Communication: Model, Analysis, and Optimization," to be published.

\bibitem{YaoJ243} 
J. Yao, {\em et al.},  ``FAS vs. ARIS: Which is more important for FAS-ARIS communication systems?," arXiv preprint \url{arXiv:2408.09067}, Aug. 2024. 

\bibitem{Espinosa24} 
P. Ramirez-Espinosa, D. Morales-Jimenez and K. K. Wong, ``A new spatial block-correlation model for fluid antenna systems," \emph{IEEE Trans. Wireless Commun.}, Early access, doi: 10.1109/TWC.2024.3434509.


\bibitem{BariahL21} L. Bariah, S. Muhaidat, P. C. Sofotasios, F. E. Bouanani, O. A. Dobre and W. Hamouda, ``Large Intelligent Surface-Assisted Nonorthogonal Multiple Access for 6G Networks: Performance Analysis," \emph{IEEE Internet Things J.}, vol. 8, no. 7, pp. 5129-5140, Apr. 2021.

\bibitem{LiS22} S. Li, L. Bariah, S. Muhaidat, A. Wang and J. Liang, ``Outage Analysis of NOMA-Enabled Backscatter Communications with Intelligent Reflecting Surfaces," \emph{IEEE Internet Things J.}, vol. 9, no. 16, pp. 15390-15400,  Aug. 2022.

\bibitem{FengH24}
 H. Feng, Y. Xu and Y. Zhao, ``Deep Learning-Based Joint Channel Estimation and CSI Feedback for RIS-Assisted Communicationss," \emph{IEEE Commun. Lett.}, vol. 28, no. 8, pp. 1860-1864, Aug. 2024. 
  
 \bibitem{ZH20}
Z. -Q. He and X. Yuan, ``Cascaded Channel Estimation for Large Intelligent Metasurface Assisted Massive MIMO,"  \emph{IEEE Wireless Commun. Lett.}, vol. 9, no. 2, pp. 210--214, Feb. 2020.

\bibitem{Aulin79} 
T. Aulin, ``A modified model for the fading signal at a mobile radio channel," \emph{IEEE Trans. Veh. Technol.}, vol. 28, no. 3, pp. 182--203, 1979.


\end{thebibliography}
\end{document}